\documentclass[preprint, preprintnumbers,amsmath,amssymb]{revtex4}

\usepackage{graphicx}
\usepackage{dcolumn}
\usepackage{bm}
\usepackage{epsfig}
\usepackage{amsfonts}

\usepackage[latin1]{inputenc} 
\usepackage{amssymb}
\begin{document}
\newcommand{\be}{\begin{equation}}
\newcommand{\ee}{\end{equation}}
\newcommand{\bea}{\begin{eqnarray}}
\newcommand{\eea}{\end{eqnarray}}
\newcommand{\n}{\nonumber\\}
\newcommand{\benu}{\begin{enumerate}}
\newcommand{\eenu}{\end{enumerate}}
\newcommand{\bite}{\begin{itemize}}
\newcommand{\eite}{\end{itemize}}

%


\title{
Atmospheric turbulence and superstatistics}

\author{C. Beck}
\affiliation{School of Mathematical Sciences, Queen Mary,
University of London, Mile End Road, London E1 4NS, UK}

\author{E.G.D. Cohen}
\affiliation{The Rockefeller University, 1230 York Avenue, New
York, New York 10021, USA}

\author{S. Rizzo}
\affiliation{ENAV S.p.A, U.A.A.V. Firenze, Italy}

\date{\today}


\maketitle

In equilibrium statistical mechanics, the inverse temperature
$\beta$ is a constant system parameter---but many nonequilibrium
systems actually exhibit spatial or temporal temperature
fluctuations on a rather large scale. Think, for example, of the
weather: It is unlikely that the temperature in London, New York,
and Firenze is the same at the same time. There are
spatio-temporal temperature fluctuations on a rather large scale,
though locally equilibrium statistical mechanics with a given
fixed temperature is certainly valid. A traveller who frequently
travels between the three cities sees a `mixture' of canonical
ensembles corresponding to different local temperatures. Such type
of macroscopic inhomogenities of an intensive parameter occur not
only for the weather but for many other driven nonequilibrium
systems as well. There are often certain regions where some system
parameter has a rather constant value, which then differs
completely from that in another spatial region. In general the
fluctuating parameter need not be the inverse temperature but can
be any relevant system parameter. In turbulent flows, for example,
a very relevant system parameter is the local energy dissipation
rate $\epsilon$, which, according to Kolmogorov's theory of 1962
\cite{K62}, exhibits spatio-temporal fluctuations on all kinds of
scales. Nonequilibrium phenomena with macroscopic inhomogenities
of an intensive parameter can often be effectively described by a
concept recently introduced as `superstatistics'
\cite{beck-cohen}. This concept is quite general and has 
been successfully applied to a variety of systems, such as
hydrodynamic turbulence,
atmospheric turbulence,
pattern formation in Rayleigh-Benard flows, cosmic
ray statistics, solar flares, networks,
and models of share price evolution \cite{erice}.
For a particular probability
distribution of large-scale fluctuations of the relevant system
parameter, namely the Gamma-distribution, the corresponding
superstatistics reduces to Tsallis statistics \cite{tsallis},
thus reproducing the generalized canonical
distributions of
nonextensive statistical mechanics by a plausible physical
mechanism based on fluctuations.

In this article we want to illustrate the general concepts
of superstatistics by a recent
example: atmospheric turbulence.
Rizzo and Rapisarda \cite{rap1, rap2}
analysed the statistical properties of turbulent wind velocity
fluctuations at Florence Airport. The data were recorded
by two head anemometers A and B on two poles 10 m high a distance
900 m apart at a sampling frequency of 5 minutes. Components of
spatial wind velocity differences at the two anemometers A and B
as well as of temporal wind velocity differences at A were
investigated.

Analysing these data, two well separated time scales can be
distinguished. On the one hand, the temporal velocity difference
$u(t)=v(t+\delta)-v(t)$ (as well as the spatial one) fluctuates on
the rather short time scale $\tau$ (see Fig.~1). On the other
hand, we may also look at a measure of the average activity of
the wind bursts in a given longer time interval, say 1 hour,
where the signal behaves approximately in a Gaussian way. The
variance of the signal $u(t)$ during that time interval is given
by $\sigma^2 = \langle u^2 \rangle -\langle u \rangle^2$, where $\langle ...
\rangle$ means taking the average over the given time interval.
We then define a parameter $\beta (t)$ by the inverse of this
local variance (i.e. $\beta=1/\sigma^2$). $\beta$ depends on time $t$, but in a much slower
way than the original signal. Both signals are displayed in
Fig.~1. One clearly recognizes that the typical time scale $T$ on
which $\beta$ changes is much larger than the typical time scale
$\tau$ where the velocity (or velocity difference) changes.

\begin{figure} 
\centerline{\epsfxsize=3.8in\epsfbox{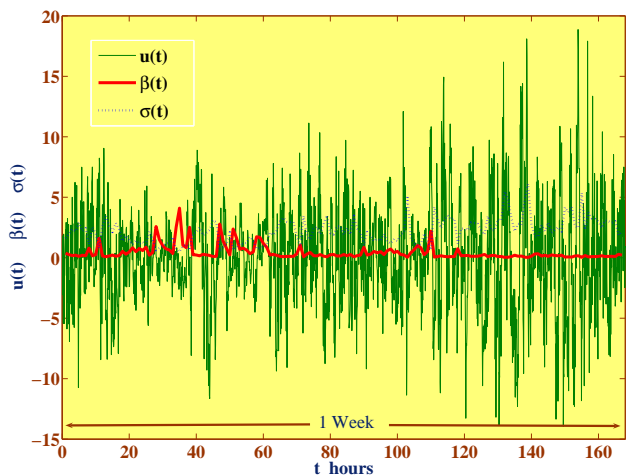}}
\caption{Time series of a temporal wind velocity difference $u(t)$
($\delta$ = 60 min) recorded by anemometer A every 5 min for one
week (green line) and the corresponding parameter $\beta(t)$ (red
line), as well as the corresponding standard deviation $\sigma(t)$
(blue dotted line), both for a 1 hour window.}
\end{figure}

Dividing the wind flow region between A and B into spatial cells,
so that air flows from one cell to another, one assumes that each
cell is characterized by a different value of the local variance
parameter $\beta$, which plays a similar role as the inverse
temperature in Brownian motion and fluctuates on the relatively
long spatio-temporal scale $T$. As mentioned before, one can then
distinguish two well separated time scales for the wind through
the cells: a short time scale $\tau$ which allows velocity
differences $u$ to come to local equilibrium described by local
Gaussians $\sim \exp[-\beta \frac{1}{2}u^2]$, and a long time
scale $T$, which characterizes the long time secular fluctuations
of $\beta$ over many cells. Similar fluctuations of a local
variance parameter are also observed in financial time series,
e.g.\ for share price indices, and come under the heading
`volatility fluctuations' \cite{bouchard}.

A terrestrial example would be a Brownian particle of mass $m$
moving from cell to cell in an inhomogeneous fluid environment
characterized by an inverse temperature $\beta$ which varies
slowly from cell to cell. The two time scales are then the short
local time scale $\tau$ on which the Brownian particle reaches
local equilibrium and a long global time scale over which $\beta$
changes significantly.  If the particle moves for a sufficiently
long time through the fluid then it samples, in the cells it
passes through, values of $\beta$ distributed according to a
probability density function
$f(\beta)$, which leads to a resulting long-term probability
distribution $p(v)$ to find the Brownian particle in the fluid
with velocity $v$ given by $p(v) \sim \int
e^{-\frac{\beta}{2}mv^2} f(\beta) d\beta$. This is like a
superposition of two statistics in the sense that $p(v)$ is given
by an integral over local statistics given by the local
equilibrium Boltzmann statistics convoluted with the statistics
$f(\beta)$ of the $\beta$ occurring in the Boltzmann statistics.
In other words, it is a `statistics of a statistics' or a
superstatistics.

Returning now to the atmospheric experiment, it is this
superstatistics which is employed here to analyse the wind data.
However, there is a fundamental difference in the interpretation
of the corresponding variables: First of all, the variable $v$
(the velocity of the Brownian particle) corresponds to the
longitudinal velocity {\em difference} in the flow (either spatial
or temporal), not the velocity itself. Secondly, since we are
analyzing turbulent velocity fluctuations and not thermal ones,
the parameter $\beta$ is a local variance parameter of the
macroscopic turbulent fluctuations and hence it does not have the
physical meaning of an inverse temperature as given by the actual
temperature at the airport. Rather, it is much more
related to a suitable power of the local energy dissipation rate
$\epsilon$.

The fluctuations of the variance parameter $\beta$ 
can be analysed
using time windows of different lengths. Rizzo and Rapisarda
carried this out for two time series of interest: for the temporal
fluctuations of the wind velocity component (in the
$x$-direction) as recorded at the anemometer A and also the
spatial fluctuations as given by the longitudinal wind velocity
differences between the anemometers A and B. The probability
distribution of $\beta$ as obtained for the temporal case is shown
in Fig.~2 for a time window of 1 hour. For comparison, the dashed
(blue) line shows a Gamma or $\chi^2$-distribution function, which
is of the general form $f(\beta) \sim \beta^{c-1} e^{-\beta/b}$,
with $b$ and $c$ appropriate constants. The
solid (red) line represents a lognormal distribution function
which is of the general form $f(\beta)\sim(1/\beta s)\exp[-(\log
(\beta/\mu))^2/(2s^2)]$, with $\mu$ and $s$ appropriate constants.
Apparently, the data are reasonably well fitted by a lognormal distribution 
(note that a different conclusion was reached by
Rizzo and Rapisarda in
\cite{rap1, rap2}).
We see that our result for atmospheric turbulence 
is similar to laboratory turbulence experiments on much
smaller space and time scales, such as a turbulent Taylor-Couette
flow as generated by two rotating cylinders.
For Taylor-Couette flow it has been shown
\cite{BCS}
that $\beta$ is indeed lognormally distributed, see
Fig.~3.

\begin{figure} 
\centerline{\epsfxsize=3.8in\epsfbox{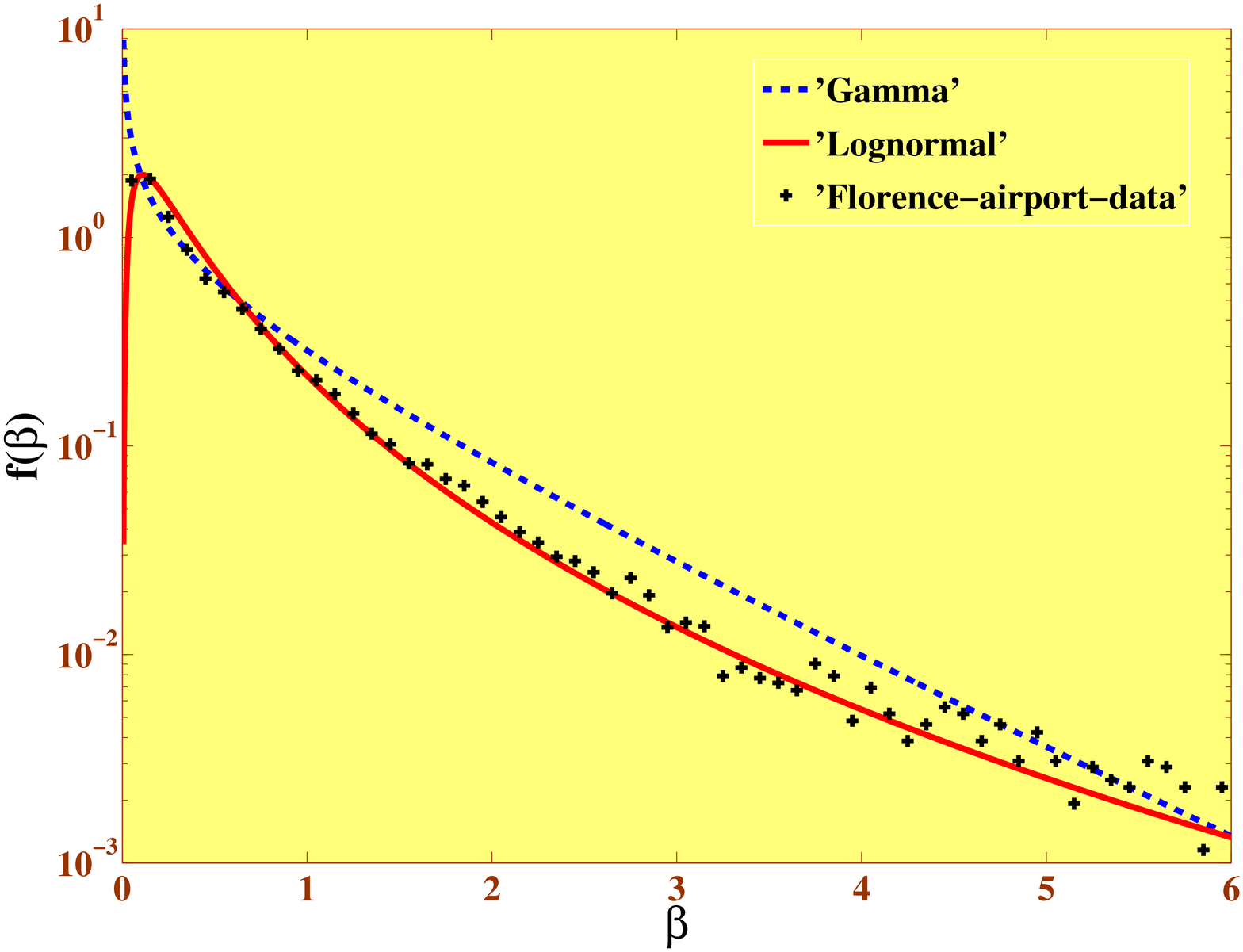}}
\caption{Rescaled probability density of the fluctuating parameter
$\beta$, as obtained for the Florence airport data. 
Also shown is a Gamma distribution (dashed blue line)
and a lognormal distribution (solid red line) sharing the same
mean and variance as the data. The data are reasonably well fitted by the
lognormal distribution.}
\end{figure}

\begin{figure}
\centerline{\epsfxsize=3.8in\epsfbox{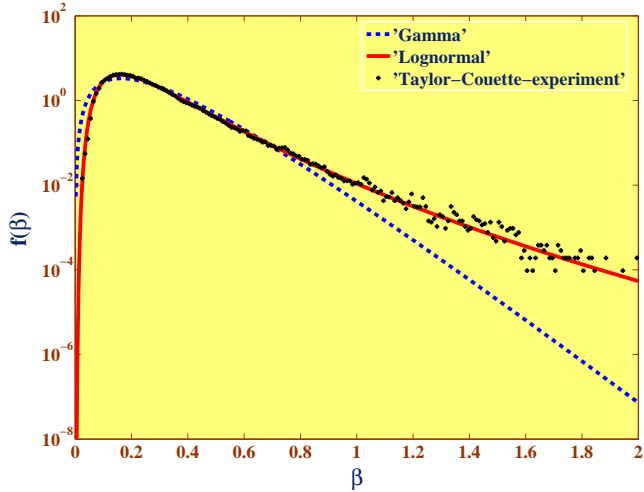}}
\caption{Probability density of the local variance parameter
$\beta$ as extracted from a time series of longitudinal velocity
differences measured in a turbulent Taylor-Couette flow at
Reynolds number $ Re=540000$ \cite{BCS}. 
A lognormal distribution yields a good fit.
} 
\end{figure}

In general, for a given nonequilibrium
system the probability density of the parameter $\beta$ 
is ultimately determined
by the underlying spatio-temporal dynamics of the system under
consideration.

The Gamma distribution results if $\beta$
can be represented by a
{\underline{sum}} of $n$ independent squared Gaussian random
variables $X^2_i$ (with $i=1,....,n)$ with mean zero, i.e.\
$\beta=\sum^n_{i=1} X^2_i>0$. The constants $c$ and $b$ above are
related to $n$.

The lognormal distribution results if $\beta$ is due to a
multiplicative cascade process, i.e.\ if it can be represented by
a {\underline{product}} of $n$ independent positive random
variables $\xi_i$, i.e.\ $\beta=\Pi^n_{i=1} \xi_i$ or $\log
\beta=\sum^n_{i=1} \log \xi_i$. Due the Central Limit Theorem,
under suitable rescaling the latter sum will become Gaussian for
large $n$. But if $\log \beta$ is Gaussian this means that
$\beta$ is lognormally distributed.

We notice that the difference between the Gamma distribution and
the lognormal distribution is essentially that of an additive
versus a multiplicative definition of $\beta$. So far there is no
theory of turbulence, but following Kolmogorov \cite{K62}, the
mechanism of the turbulent motion of the fluid is critically
determined by the transfer mechanism of the energy dissipation
between neighboring cells and between different spatial scales in
the flow. A multiplicative cascade process is expected to
lead to a lognormally distributed $\beta$.
It seems that the above mentioned transfer mechanism for
energy dissipation is similar for turbulent wind
fluctuations and laboratory turbulence, which are performed under
very different conditions. The spatial scale of environmental
turbulence as measured at the airport is much larger than in the
laboratory, moreover the Reynolds number fluctuates for the wind
measurements, whereas in the laboratory experiments it is
controlled.


The probability density $p(u)$ of longitudinal wind velocity
differences $u$ (either temporal or spatial) as measured at the
airport has strong deviations from a Gaussian distribution
and it exhibits prominent ('fat') tails 
(see Fig.~4). In superstatistical models one can
understand these tails simply from a superposition of Gaussian
distributions whose inverse variance $\beta$ fluctuates on
a rather large spatio-temporal scale. In the long-term run one has
$p(u) \sim \int_0^\infty f(\beta) 
e^{-\frac{1}{2}\beta u^2} d\beta$,
and generically these types of distributions $p(u)$ exhibit broader
tails than a Gaussian.

\begin{figure}
\centerline{\epsfxsize=3.8in\epsfbox{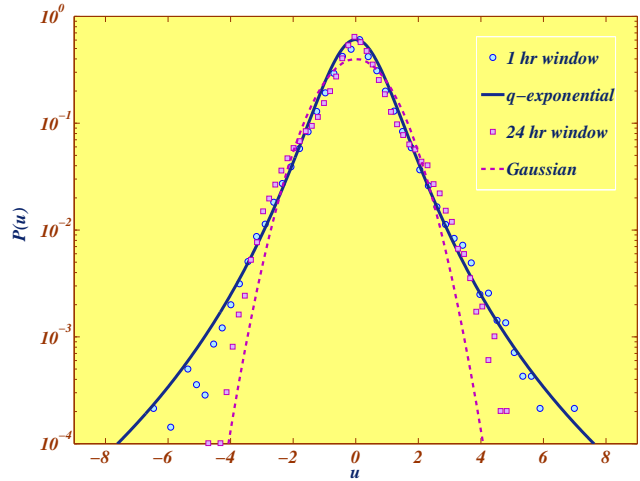}}
\caption{Probability density $p(u)$ of temporal
wind velocity differences as observed at the airport.
}
\end{figure}

For the special case that $f(\beta)$ is a Gamma distribution
the integral can be explicitely evaluated, and one ends up
with the generalized canonical distributions ($q$-exponentials)
of nonextensive statistical mechanics, i.e.\
$p(u) \sim
(1+\tilde{\beta} (q-1)\frac{1}{2}u^2)^{-\frac{1}{q-1}}$, where
$\tilde{\beta}$ is proportional to the average of $\beta$
and $q$ is an entropic parameter
\cite{beck-cohen, tsallis}. These distributions asymptotically
decay with a power law.
For other $f(\beta)$ (such as the lognormal
distributions relevant in our case), the integral
cannot
be evaluated explicitly, and more complicated
behaviour arises. However, it can be shown that for
sharply peaked distributions $f(\beta)$ a $q$-exponential
for $p(u)$ is often a good approximation 
provided $|u|$ is not too large \cite{beck-cohen}.

Quite generally, the superstatistics approach also
gives a plausible physical interpretation to the entropic index
$q$. One may generally define  
\begin{equation}
q=\frac{\langle \beta^2 \rangle}{\langle \beta \rangle^2},
\label{333}
\end{equation}
where $\langle \beta \rangle =\int f(\beta) \beta d\beta$ and
$\langle \beta^2 \rangle = \int f(\beta) \beta^2 d\beta$ denote
the average and second moment of $\beta$, respectively. Clearly,
if there are no fluctuations in $\beta$ at all but $\beta$ is
fixed to a constant value $\beta_0$, one has $\langle \beta^2
\rangle = \langle \beta \rangle^2=\beta_0^2$, hence in this case
one just obtains $q=1$ and ordinary statistical mechanics
arises. On the other hand, if there are temperature
fluctuations (as in most nonequilibrium situations) then those are
effectively described by $q>1$. For the special case that
$f(\beta)$ is a Gamma-distribution, the
$q$ obtained by eq.~(\ref{333}) coincides with Tsallis' entropic
index $q$ (up to some minor correction arising from the local
$\beta$-dependent normalization constants).
But the superstatistics concept is more general in that it also
allows for other distribution $f(\beta)$, as for example the
lognormal distribution observed
in Fig.~2 and 3. 
General superstatistics can
lead to a variety of distributions $p(u)$ with prominent
('fat') tails, i.e.\ not only power laws but, for example, also
stretched exponentials tails and much more.

The atmospheric turbulence data seem roughly consistent with
Kolomogorov's general ideas of a lognormally
distributed fluctuating energy dissipation
rate, as are the laboratory turbulence data.
In that connection comparison with long range oceanic measurements of a
similar kind as the atmospheric wind experiments discussed here
might be instructive, testing yet larger scales.

\end{document}